\documentclass{aa}
\usepackage{babel,amssymb,amsmath,graphicx,hyperref}

\begin{document}

\title{Shadow of rotating black holes on a standard background screen}

   \author{S.V.Repin \inst{1}
          \and
          D.A.Kompaneets \inst{1}
          \and
          I.D.Novikov \inst{1,2,3}
          \and
          V.A.Mityagina \inst{4,5}
          }

   \institute{Astro Space Center, Lebedev Physical Institute of RAS,
               84/32, Profsoyuznaya str., 117997, Moscow, Russia
         \and
             The Niels Bohr International Academy, The Niels Bohr
             Institute, Blegdamsvej 17, DK-2100, Copenhagen, Denmark
         \and
             National Research Center Kurchatov Institute,
             1, Kurchatova sq., 123182, Moscow, Russia
         \and
             Information Technology Lyceum, No.1533,
              16, Lomonosovsky av., Moscow, Russia \\
              \email{info@lit.msu.ru}
          \and
             Moscow State University,
             GSP-1, Leninskie Gory, 119991, Moscow, Russia
              }

\abstract{\large
     We present the shape of the black hole shadow on the standard
background screen as it is registered by the distant observer. The
screen is an infinite plane, emitting the quanta uniformly
distributed to a hemisphere. The source of emission is considered to
be optically thin and optically thick. It is shown that the shape of
a black hole shadow depends crucially on the angle between the plane
and the view line to the distant observer. The shadow shapes for the
different values of this angle are also presented. Both
Schwarzschild and Kerr metrics are considered.}

\keywords{Black hole physics; Gravitational lensing: strong;
Methods: numerical}

\maketitle

\section{Introduction}

      Black holes are the most intriguing objects in astrophysics.
Their existence was predicted by the General theory of Relativity
(GR). However, the radii of black holes appear to be so small that
they cannot be observed immediately yet. Usually, observers register
the effects in the accretion disks surrounding the black holes, but
not the black holes themselves. One possible way to observe black
holes (at least theoretically!) is to observe their shadows.

      The black hole shadow is the area in the celestial sphere
around the black hole position from which no one quanta comes to the
observer. The size of this area does not coincide with the size of
the event horizon. The aim of this paper is to consider the
structure of the shadow in addition to the many considerations
performed earlier \citep{Amarilla_2015, Ghasemi-Nodehi_2015,
Neronov_2016, Abdujabbarov_2016, Cunha_2016, Shipley_2016}. The
largest angular size is the black hole shadow in the center of our
Galaxy and it is approximately 54 micro arc seconds in diameter
\citep{Gillessen_2009} under the assumption that the black hole is
described by the Schwarzschild metric and the standard background
screen is perpendicular to the view line. The second largest is the
elliptic galaxy M87 ($\theta\approx 22$ to 41 micro arc seconds
\citep{Walsh_2013}) and the third is the Andromeda galaxy M31
($\theta\approx 14$ to 30 micro arc seconds \citep{Bender_2005}). It
is evident that the objects of the microsecond size can be observed
in detail and investigated only by the inter\-fe\-ro\-meters because
the angular resolution of the radiotelescopes is much lower. The
Event Horizon Telescope (EHT) could be one of these instruments
\citep{Akiyama_2015} as well as Millimetron in Russia
\citep{Kardashev_2014} and GRAVITY in Chile
\citep{Lacour_2014,Vincent_2014}.

      The shape of a black hole shadow depends on the objects and
the matter that presents in the vicinity of the black hole. The
image of the black hole shadows in M87 \citep{Dexter_2012,
Moscibrodzka_2016} and Sgr~A$^*$ \citep{Broderick_2016,
Johannsen_2016} have been simulated many times under the different
assumptions. The si\-mu\-la\-tions take into account the jet and the
accretion disk, the effects of hydrodynamics, magnetic fields,
optical depth of the disk and a lot of other effects. As a result,
these simulations can give a detailed image of a particular object,
but cannot reveal the effect of pure lensing. But sometimes, the
general view is necessary.  So, it would be interesting to
demonstrate the effects of General Relativity only, i.e. to present
a shadow of a black hole on the sample background of a very simple
structure like a bright infinite plane (we will call it a standard
screen) and reveal the details of the image, which can be directly
observed and which will indicate that the object is definitely the
black hole.

     In this paper we present the shape of a black hole shadow
and the distribution of the emission intensity in the vicinity of
the edge on the background of a standard screen as it is registered
by the distant observer. The other model of the emission of the
standard screen will be considered at the end of Section 6.

\section{Geometry}

     We should consider a black hole that is placed in front of a
bright plane, which is infinitely far from the observer. We will
consider the different inclination of the standard screen to the
view line. The observer should also be at infinity with respect to
both the black hole and the bright plane. However, these assumptions
are very inconvenient for the computations and it would be desirable
to avoid the infinite values of variables. Therefore, we should
place our objects at the finite distances, but, except for the black
hole, they should be outside the area of strong gravity. That is the
model we accept.

     Let the standard screen (a bright plane) is perpendicular to
the $x$-axis and crosses this axis at the coordinate
$x_0=-1000\,r_g$, where $r_g = 2Gm/c^2$. It means that the equation
of this plane is $x=-1000\,r_g$ ($y$ and $z$ can have arbitrary
values). Later we will consider the change of the angle between the
plane and the x-axis. The observer is also on the $x$-axis, but her
coordinates are: ($1000\,r_g,0,0$). The black hole is located in the
center of the Cartesian system, so its coordinates are: ($0,0,0$).
This design means that both the standard screen and the observer are
in the non-relativistic region, where classical mechanics
approximations can be applied successfully. The region of strong
gravitation, where the General relativity should be counted,
surrounds the center of the coordinate system.

      When we change the inclination of the standard screen
(with respect to the view line) we assume that the distance between
the screen and the black hole remains unchanged and equals to
$1000\,r_g$. The same geometry is used when we consider the spinning
black hole.

\section{Simulation methods}

     To build the shadow of the black hole we need to si\-mu\-late
the trajectories of the huge amount of quanta, emitted uniformly by
a standard screen in a solid angle $2\pi$ and select only those,
which are registered by (comes to) the distant observer. This is a
direct way, but it requires an extremely long computation time
because much of the quanta have to be discarded. Because of that, it
is preferable to simulate the quanta propagation in opposite
direction, i.e. from the observer to the standard screen. Thus, we
count the quanta which are registered by the observer only and avoid
the calculation of the trajectories which pass by him. This method,
however, can be applied to the Schwarzschild black holes, but not to
the spinning ones, because the ray reversibility principal is not
satisfied in the Kerr metric. But this difficulty can be overcome by
changing the direction of rotation of the black hole to the opposite
one.

     So, one should consider the trajectories of the quanta, which
start from the observer and move to the black hole so that the
maximum value of their impact parameter would be 3-4 times higher
than the value at which the quanta are captured by the black hole.
It corresponds to the impact parameter of approximately $8\div
10\,r_g$.

     The standard screen is considered to be optically thin, so its
brightness is a constant.

     The equations of motion for the quantum can be reduced to the
system of six ordinary differential equations
\citep{Zakharov_1994,Zakh_Rep_1999, Zakh_Rep_2002}:
\begin{eqnarray}
   \frac{dt}{d\sigma}
                           & = &
      - a \left(a \sin^2\theta - \xi\right) +
      \frac{r^2 + a^2}{\Delta}
       \left(r^2 + a^2 - \xi a\right),
                        \label{eq1}                \\
   \frac{dr}{d\sigma} & = & r_1,       \label{eq2} \\
   \frac{dr_1}{d\sigma}    & = &
      2r^3 + \left(a^2 - \xi^2 - \eta\right) r +
      \left(a - \xi\right)^2 + \eta,               \\
   \frac{d\theta}{d\sigma} & = & \theta_1,         \\
   \frac{d\theta_1}{d\sigma}
                           & = &
      \cos\theta \left(\frac{\xi^2}{\sin^3\theta} -
                       a^2 \sin\theta
                 \right),              \label{eq5} \\
   \frac{d\phi}{d\sigma}   & = &
      - \left(a - \frac{\xi}{\sin^2\theta}\right) +
      \frac{a}{\Delta}
           \left(R^2 + a^2 - \xi a \right),
                     \label{eq6}
\end{eqnarray}
where $t,1/r,\theta,\phi$ are the Boyer -- Lindquist coordinates,
$\eta = Q/M^2E^2$ and $\xi = L_z/ME$~-- the Chandrasekhar constants,
$Q$~-- the Carter constant \citep{Carter_1968}, $E$~-- the quantum
energy at the infinity, $L_z$~-- the projection of the momentum of
the quantum to z-axis, $\sigma$~-- the independent variable, $\Delta
= r^{-2}-2/r+a^2$. The auxiliary variables $r_1$ and $\theta_1$ do
not have adequate physical sense, they are necessary to relieve the
computational problems only.

     The system (\ref{eq1})-(\ref{eq6}) has been solved numerically
by the program package, which applies the combination of Gear
\citep{Gear_1971} and Adams methods and is distributed freely in
Internet \citep{Petzold_1983}.

\section{Trajectories}

     We take into consideration all the trajectories, which start on
the standard screen and come to the distant observer (the direction
of the propagation we've discussed above). The vast majority of the
photons arrive at the observer after the deviation in the field of a
black hole at the angle from 0 to 90 degrees. These rays form the
outer bright ring around the shadow of a black hole; it is a kind of
radiance or a crown. The impact parameter of the quanta here is
greater than approximately $3.1\,r_g$ in Schwarzschild metric. In
other words, the value of the impact parameter separates the shadow
from the outer bright area. In Kerr metric, the situation is
asymmetric and depends on the direction of the black hole rotation.

\begin{figure}[tbh]
  \centerline{
  \includegraphics[width=4.5cm]{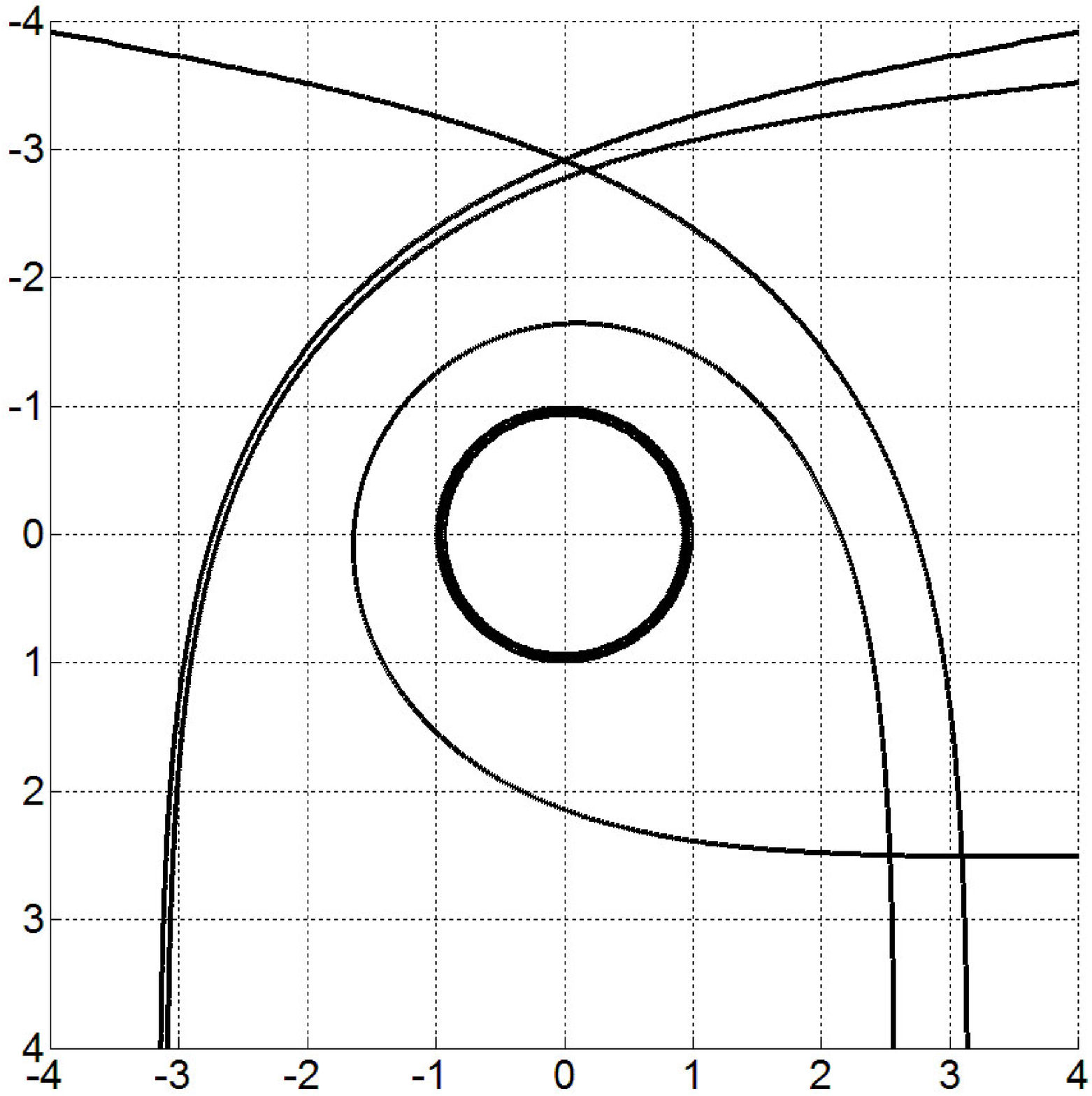}
  \includegraphics[width=4.5cm]{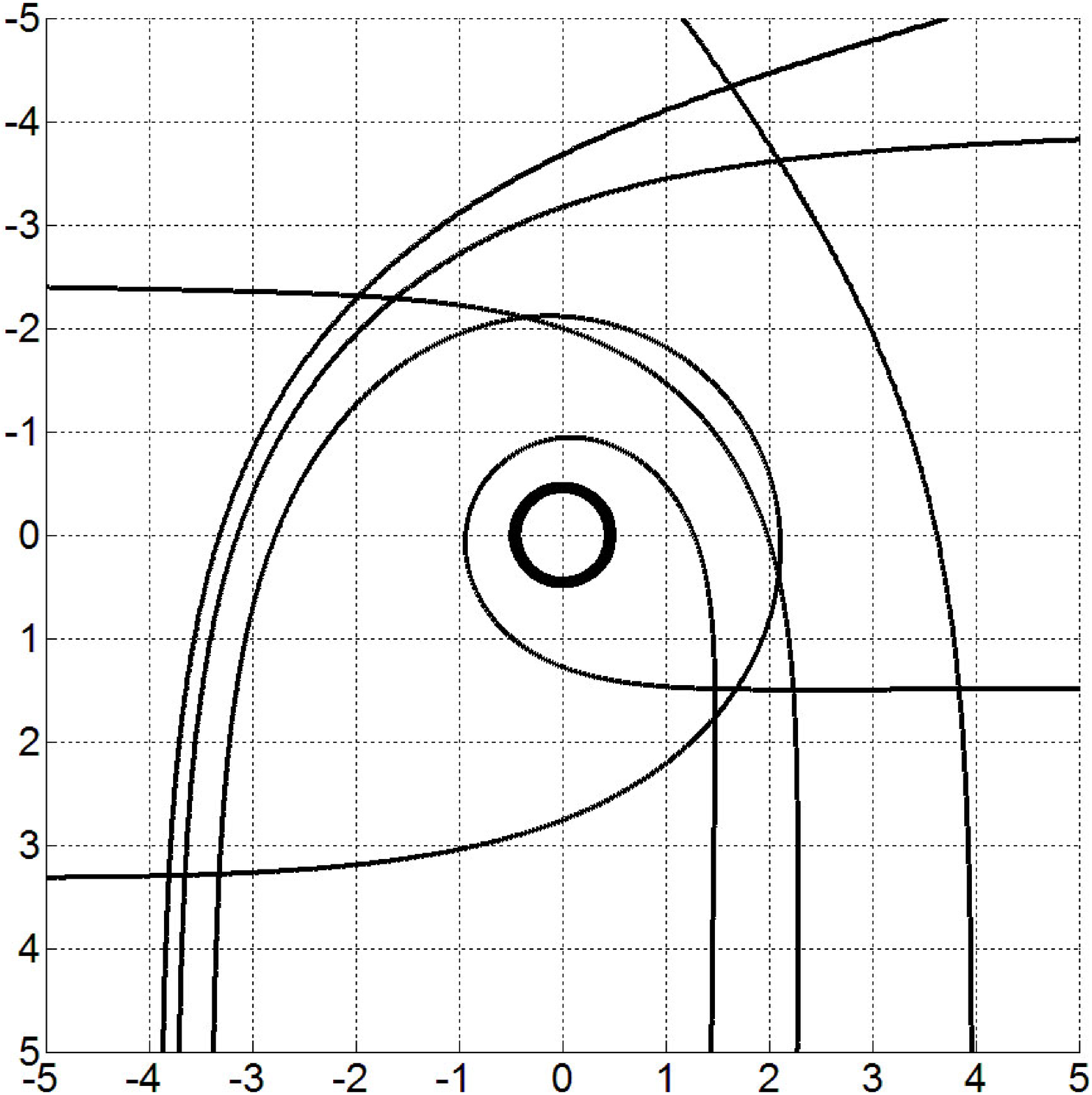}
             }
  \centerline{
  \includegraphics[width=4.5cm]{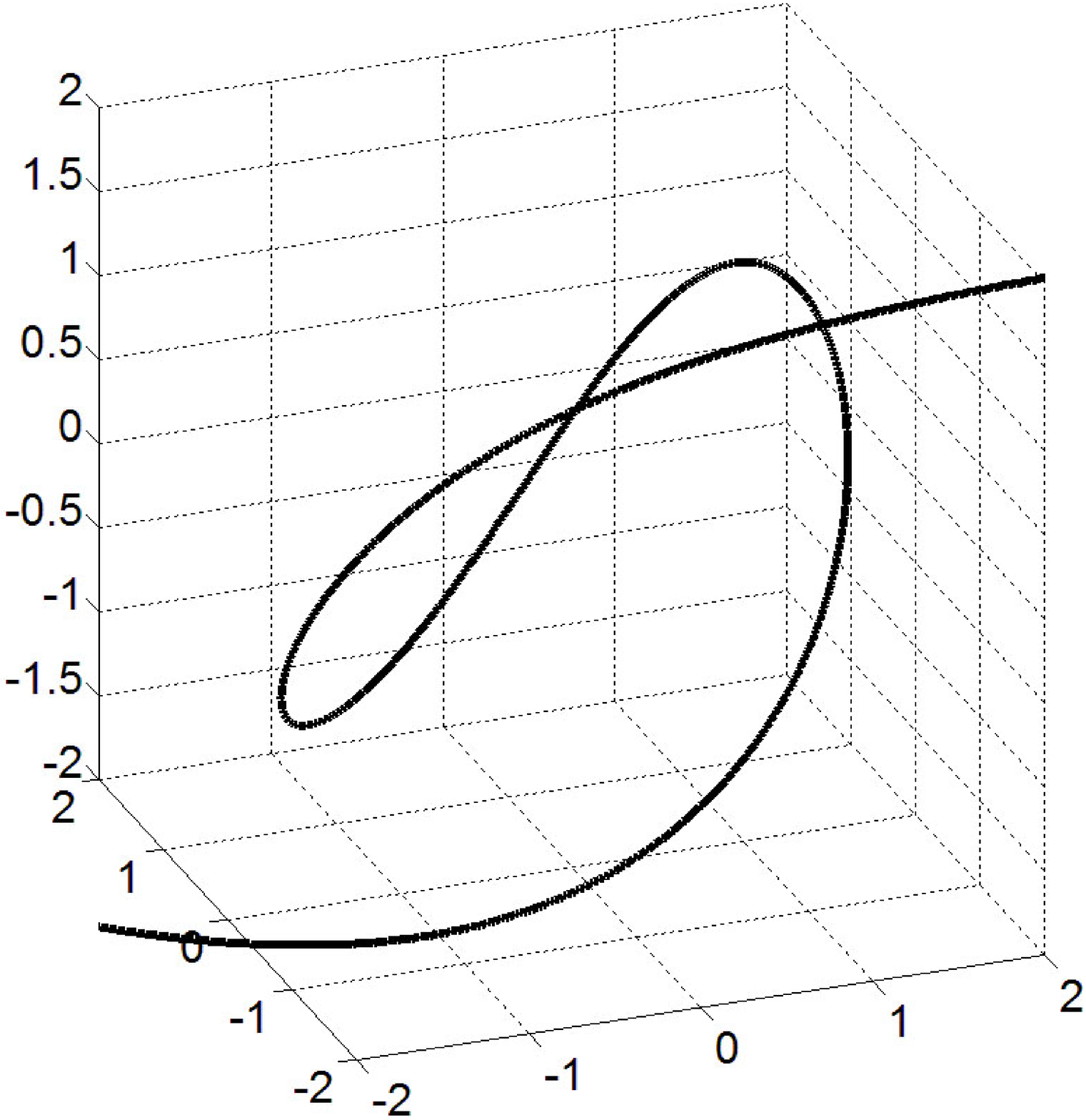}
  \includegraphics[width=4.5cm]{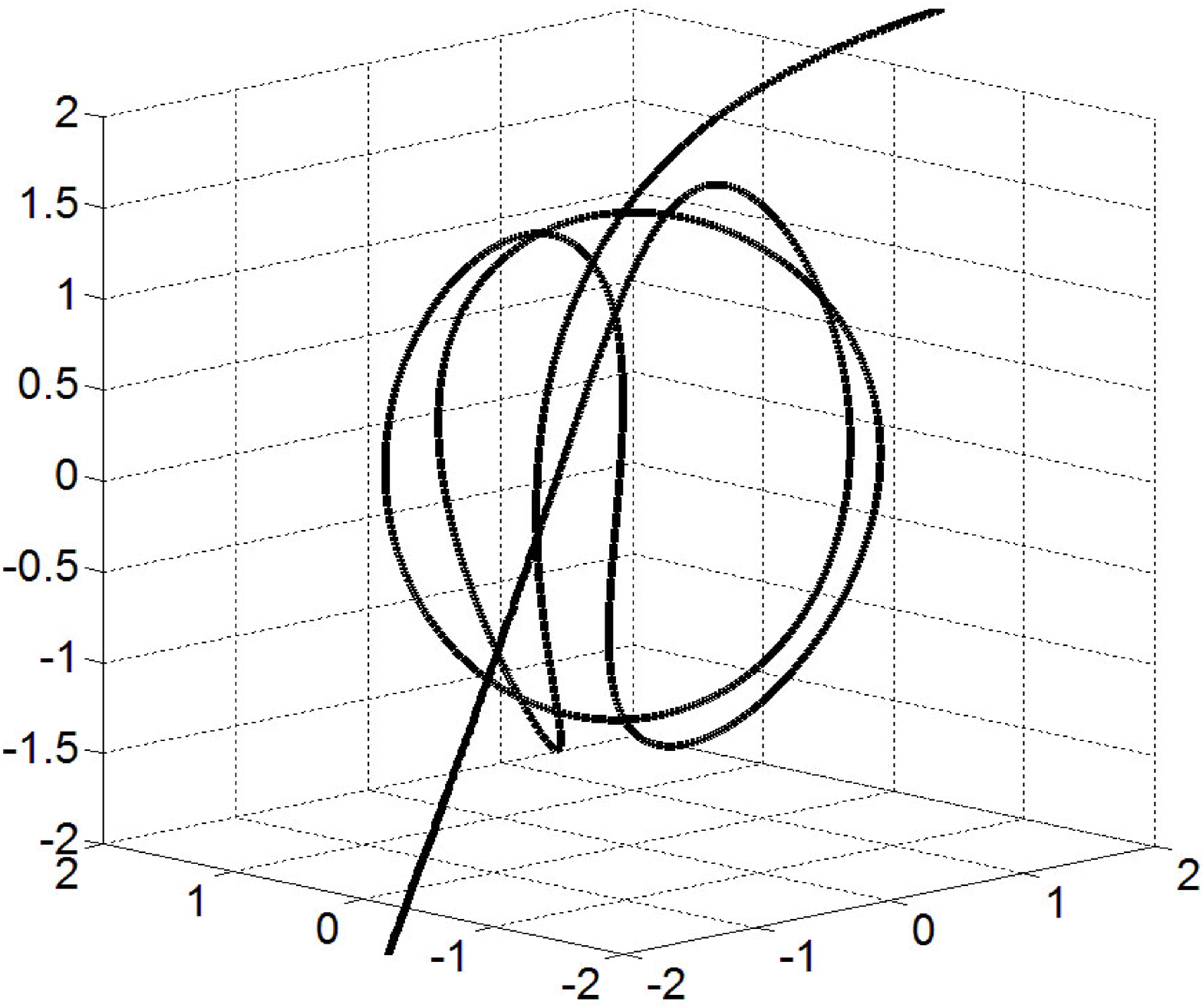}
              }
  \caption{Example of very long and complicated trajectories
           of the quantum in the Cartesian coordinate system.
           The black hole has extreme spin and it is
           placed at the point (0,0,0). The top panels represent
           the flat trajectories. See the details in the text.}
  \label{Trajectory1}
\end{figure}

      There is some quanta that performs a few revolutions (one or
more) around the black hole before they arriving at the observer.
These quanta form a thin annulus inside the shadow area. The
structure of this annulus we will discuss below.

      The top panels in Fig.~\ref{Trajectory1} demonstrate the flat
trajectories in Schwarzschild (left) and extreme Kerr (right)
fields, respectively. Both panels demonstrate: (a) the trajectories
with the same impact parameter that pass on either sides (right and
left) of the black hole; (b) the trajectories, which turn by
$90^\circ$ angle from their initial direction of propagation; (c)
the trajectories which turn by $270^\circ$ angle (3/4 of the full
revolution). The quanta of (b) section correspond to the border
between the exterior bright area and the area of the shadow. The
quanta of (c) section separate the dark area of the shadow and outer
edge of the bright annulus inside the shadow. In Kerr metric the
left and right trajectories are asymmetric because one of the quanta
flies in the same direction as the spinning of the black hole (and
the spinning of the inertial frame), while the other one flies
counter to the spinning.

      In Kerr metric the trajectories can be very unusual and
confusing, especially the non-flat trajectories. The bottom panels
in Fig.~\ref{Trajectory1} demonstrate two of these trajectories that
make one and three revolutions around the black hole. All these
complicated trajectories make the contribution to the image of the
shadow and to the brightness of the thin annulus in the shadow area
and should be taken into consideration. Thus, during the simulation
process we have found the trajectory with 29 revolutions and a lot
of trajectories with more than 5 revolutions.

\section{Black hole shadows}

\subsection{Schwarzschild black hole}

     As it has already been mentioned that a shadow of a black hole
is an area of the celestial sphere from which no one quanta arrives
to the observer.

\begin{figure}[tbh]
  \centerline{
  \includegraphics[width=4.4cm]{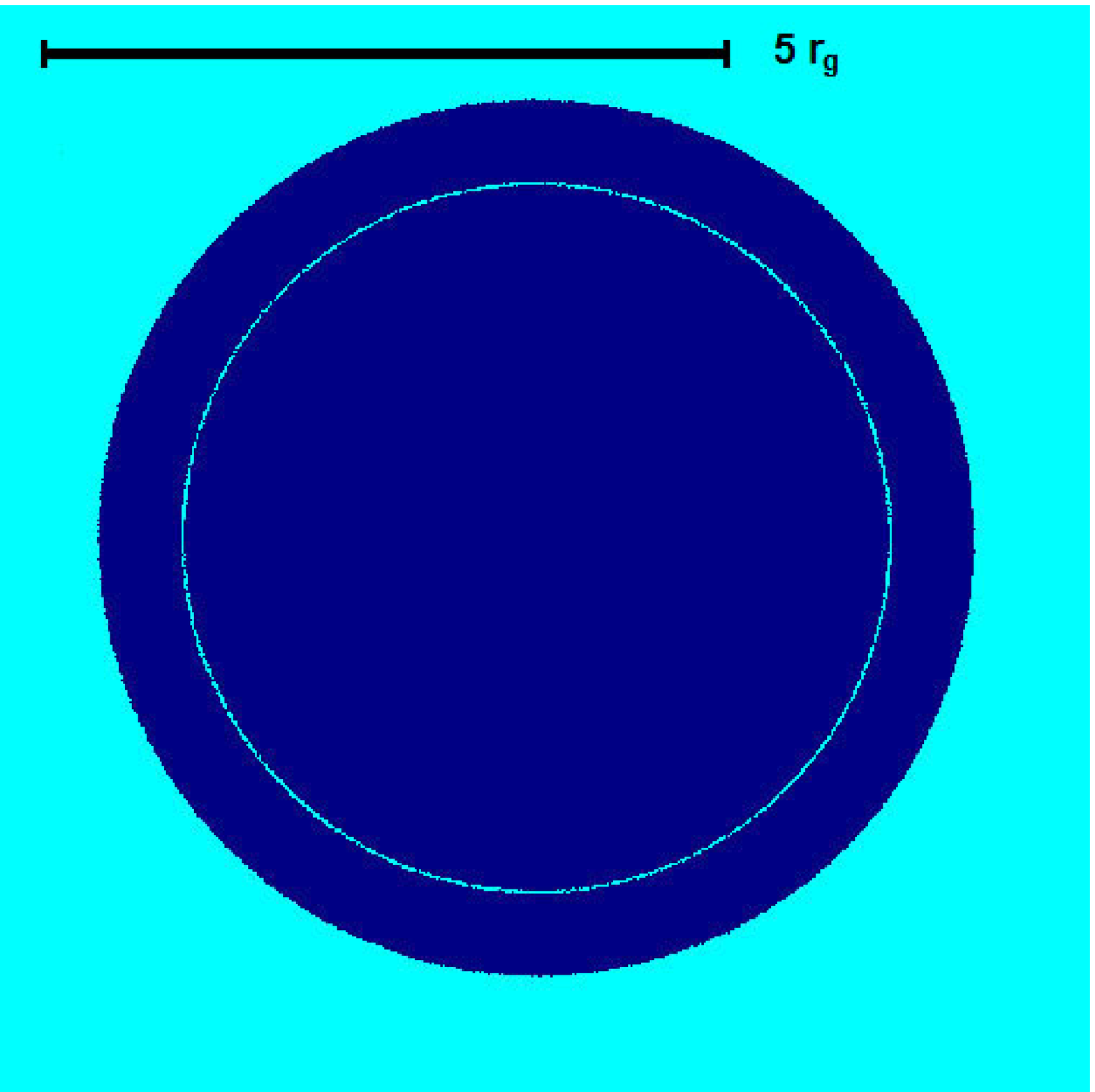}
  \includegraphics[width=4.4cm]{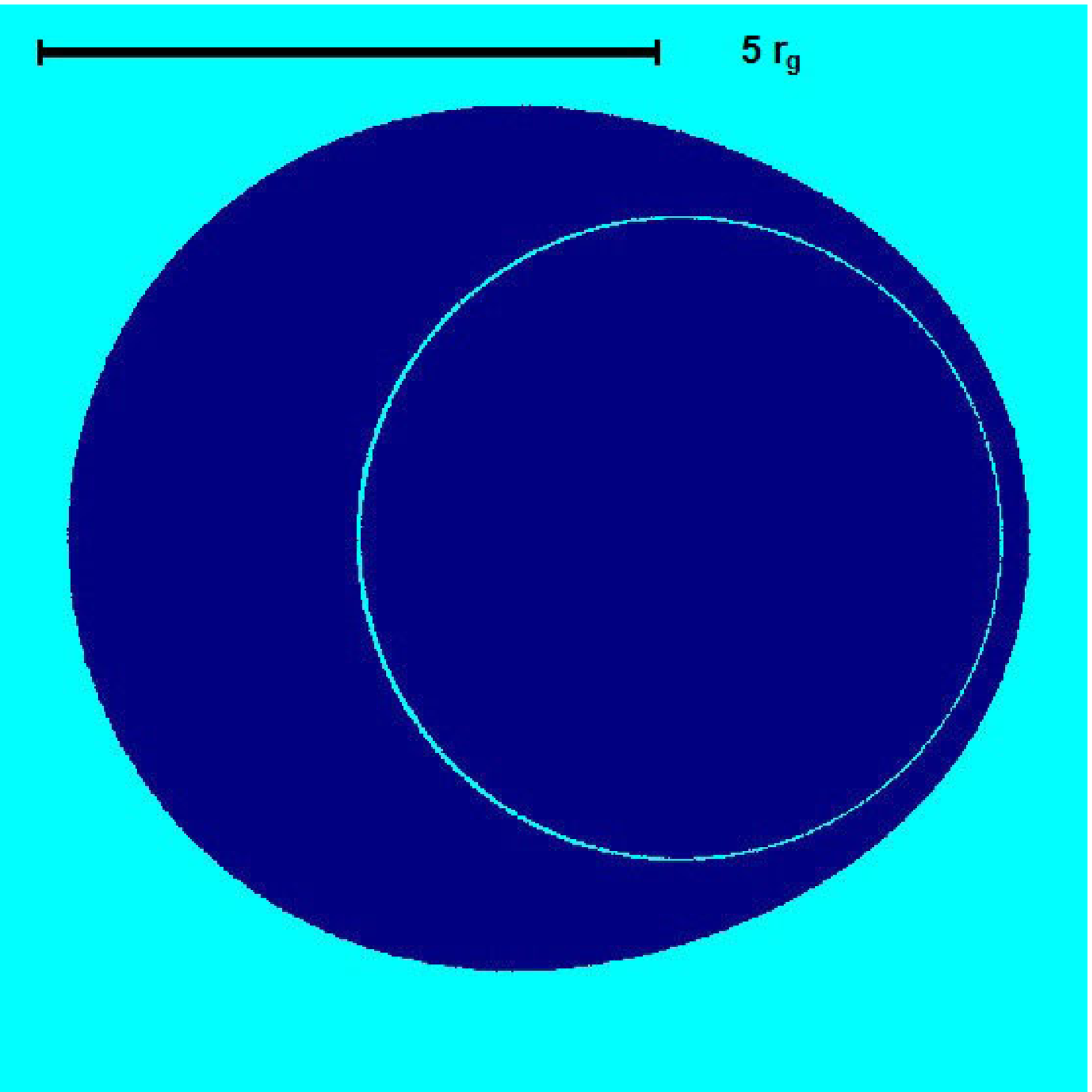}
              }
  \centerline{
  \includegraphics[width=4.4cm]{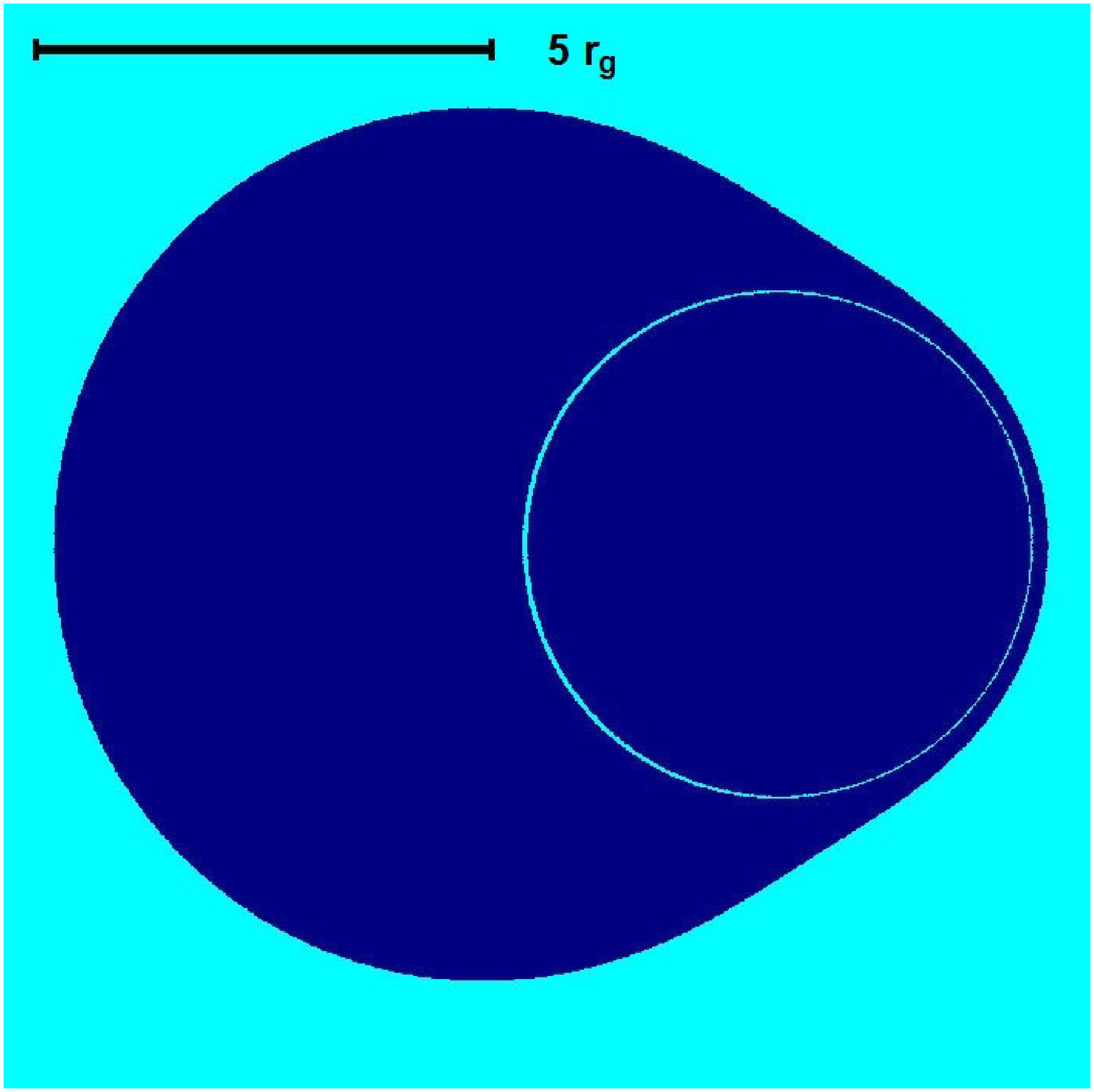}
  \includegraphics[width=4.4cm]{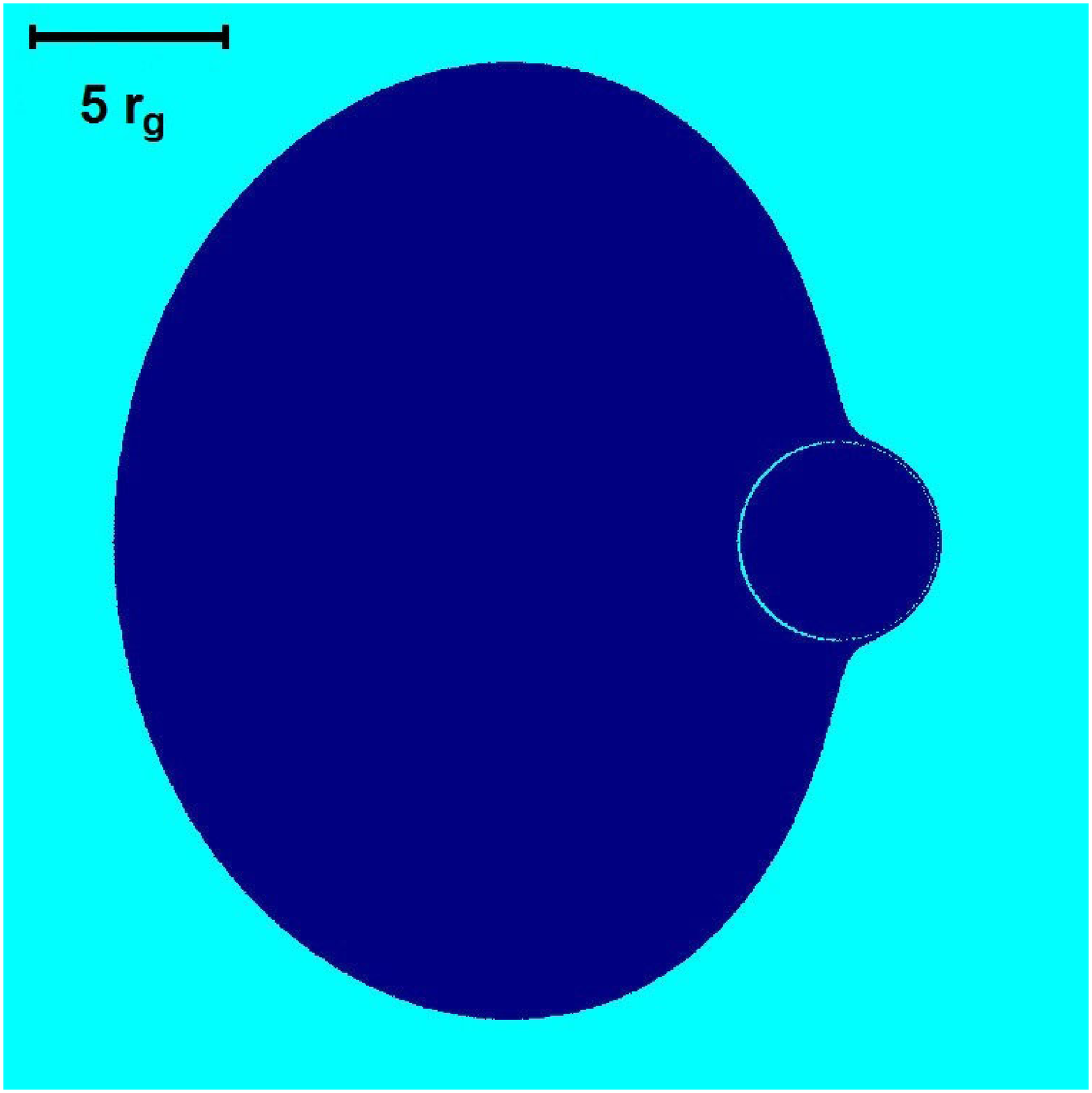}
              }
  \caption{Shadows of Schwarzschild black hole with different
           inclination angle of the standard screen. From top left
           panel to bottom right the angles are: $0^\circ\!$,
           $45^\circ\!$, $60^\circ$ and $81^\circ$. The segments in
           the top left corners indicate the scale.}
  \label{Schw_shadow1}
\end{figure}

      The shape of the shadow of the Schwarzschild black hole shown
in Fig.~\ref{Schw_shadow1}. The top left panel represents the model
where the standard screen is perpendicular to the line of view. The
inner edge of the internal annulus should have the radius
$r={3\sqrt{3}/2 \cdot r_g} \approx 2.598\,r_g$ \citep{Zel_Nov_1964},
that corresponds to the impact parameter at which the photon is
captured by the black hole. Our numerical simulation confirms the
fact that the internal radius of the annulus equals to $2.598\,r_g$
as well as the fact that the outer radius of the annulus is greater
and equals to $2.614\,r_g$. Between these values there is an
infinite number of very thin annuli, corresponding to the increasing
number of the revolutions of quantum with decreasing the impact
radius from its largest value of $2.614\,r_g$ to the lowest one. It
means that the thin annulus in the shadow in reality consists of a
set of very thin rings and each of these thin rings includes the
quanta that executes the same number of revolutions around the black
hole. The outer radius of the shadow area is approximately
$3.084\,r_g$.

     The top right panel in Fig.~\ref{Schw_shadow1} demonstrates the
shadow of the Schwarzschild black hole when the angle between the
standard screen and the view line equals to $45^\circ$. The image
here becomes asymmetrical and the shadow area increases.
Nevertheless, the inner annulus has approximately the same size.
This fact is clear from general physics because the annulus includes
the quanta that make one or more revolutions around the black hole,
so their impact parameter should be approximately unchanged.

     The bottom panels in Fig.~\ref{Schw_shadow1} presents the
shadow of the black hole when the angle between the screen and the
view line equals to $60^\circ$ and  $81^\circ$. The size of the
shadow grows with increasing the inclination angle and at $81^\circ$
its height becomes greater than its width whereas the size of the
inner annulus remains essentially without changes.

\subsection{Spinning black hole}

\begin{figure}[tbh]
  \centerline{
  \includegraphics[width=4.7cm]{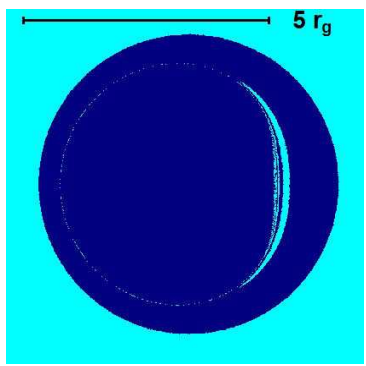}
             }
  \caption{Shadow of the spinning black hole.
           The view line is perpendicular to both
           the standard screen and the spin axis
           of the black hole. The segment at the top
           gives the scale. The shadow diameter (horizontal) is
           $6.11\,r_g$.}
  \label{Kerr_shadow1}
\end{figure}

     The shape of the shadow of a spinning black hole (Kerr metric)
is shown in Fig.~\ref{Kerr_shadow1}. The Figure represents the
geometry when both the standard screen and the spin axis are
perpendicular to the view line. The horizontal shadow diameter is
approximately $6.11\,r_g$. The shadow shape is slightly asymmetrical
and inside the shadow area we can see the thin annulus as it occurs
in the Schwarzschild metric. But, unlike the Schwarzschild metric,
the annulus is asymmetrical and its center does not coincide with
the center of the shadow image. Moreover, one can distinguish a
number of very thin rings on the right hand side of the inner
annulus, but in the left side these rings are indistinguishable
because the distance between them is very small. Each of these rings
refers to the quanta that executes a definite number of revolutions
around the black hole (the quanta of the outer ring execute only one
revolution).

     The area inside the annulus is known as the shadow of a black hole
under the assumption that the emission drops to the black hole from
the solid angle $4\pi$. The edge of this area refers to the impact
parameter at which the quantum drops into the black hole. The form
of the shadows in this sense of the word is well known and it is
presented in \cite{Chandrasekhar_1983} for marginally spinning black
hole. This form coincides with that shown in
Fig.~\ref{Kerr_shadow1}.

\section{Emission intensity}

     The form of the black hole shadow, presented in
Figs.~\ref{Schw_shadow1}-\ref{Kerr_shadow1} is very interesting from
a theoretical point of view. But for the observation carried out by
space telescopes, the distribution of the emission intensity around
and inside the shadow area is more important. This intensity is not
a constant, of course.

     To simulate the intensity distribution one should consider the so
called \glqq picture plane\grqq{}, which is placed just in front of
the observer. When we release a large amount of quanta in the
direction of the black hole we distribute them uniformly in the
tight solid angle and, consequently, they are distributed uniformly
in the picture plane as well (the bright region in
Figs.~\ref{Schw_shadow1}-\ref{Kerr_shadow1} has the uniform
intensity). All these quanta, when they reach the standard screen,
form there a non-uniform distribution. The emission of the standard
screen is, however, uniform, according to the model. The idea is
that we should find the continuous function which transforms the
non-uniform distribution on the standard screen to the uniform one.
Once found, this function gives us the possibility to assign each
quantum a weight coefficient. If we then count all the quanta on the
picture plane with their weight coefficients, we will have the true
distribution of the emission intensity.

\begin{figure}[tbh]
  \centerline{
  \includegraphics[width=4.4cm]{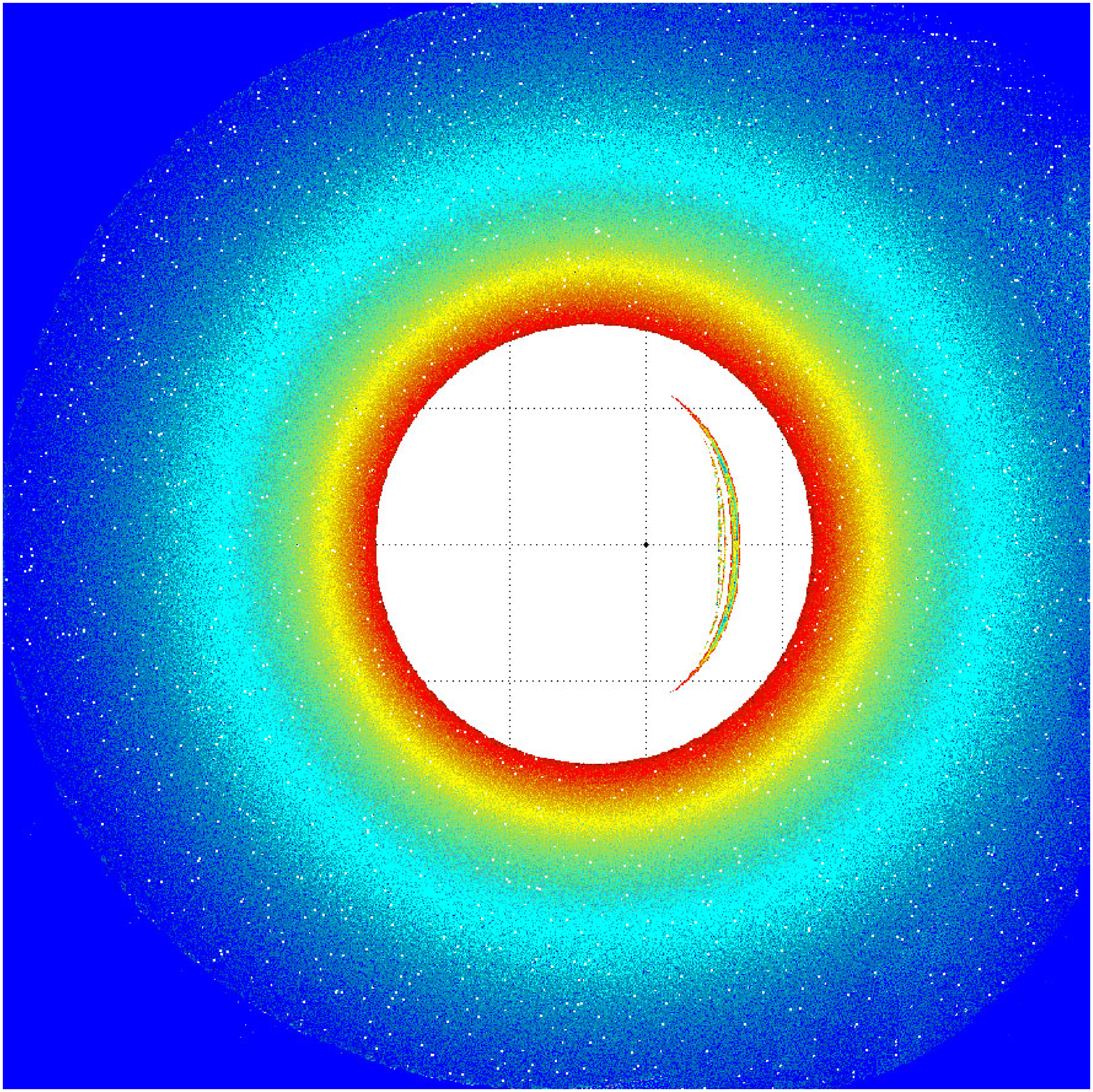}
  \includegraphics[width=4.4cm]{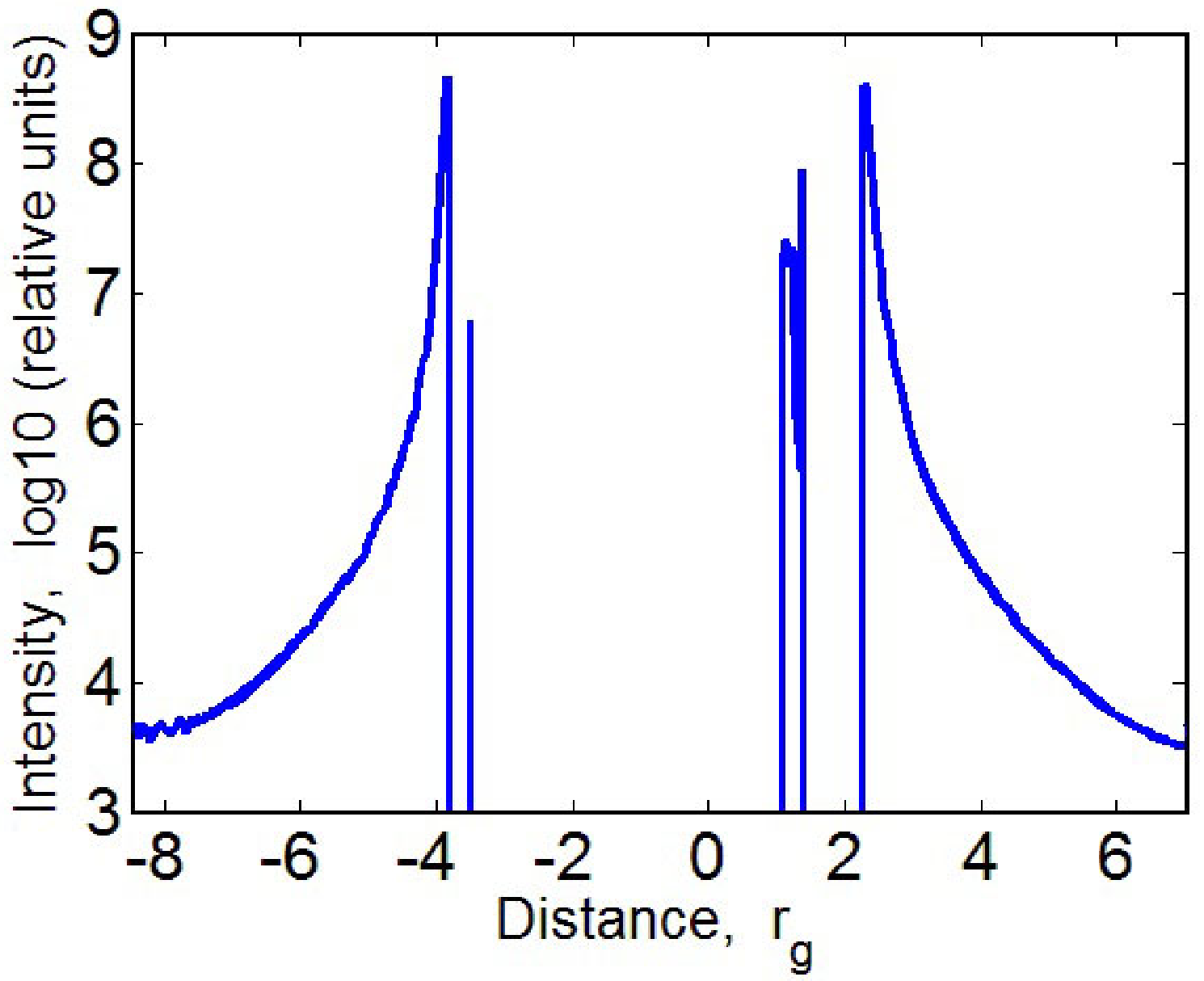}
              }
  \caption{Distribution of the emission intensity for the spinning
           black hole. The intensity is presented in conditional
           colors; the shadow area is white.
           The grid step is $2r_g$. The bold dot marks
           the center of the black hole. The right panel
           demonstrates the distribution of the intensity in
           the equatorial plane. The logarithmic scale is used
           along the $y$-axis.}
  \label{Emission_intens1}
\end{figure}

     The distribution of the emission intensity for the marginally
spinning black hole in the geometry where the view line is
perpendicular to both the standard screen and the spinning axis is
shown in Fig.~\ref{Emission_intens1}. The $y$-axis is presented in
the logarithmic scale and arbitrary (relative) units. The right
panel demonstrates the distribution of the intensity in the
equatorial plane. As it follows from the figure, the brightness of
the image increases with decreasing the radial coordinate and the
increase covers more than four orders of magnitude. In reality,
however, the intensity tends to infinity and the vertical line is
the asymptote because the brightness of the optically thin standard
screen becomes infinite at the edge. Except that, we consider in
simulation the finite number of quanta and cannot obtain as a result
the infinite intensity in principal. It means that the very top part
of the curve may lay slightly higher than it is shown in the plot.
The brightness of the inner annuli tends also to infinity at their
edges.

\begin{figure}[tbh]
  \centerline{
  \includegraphics[width=4.4cm]{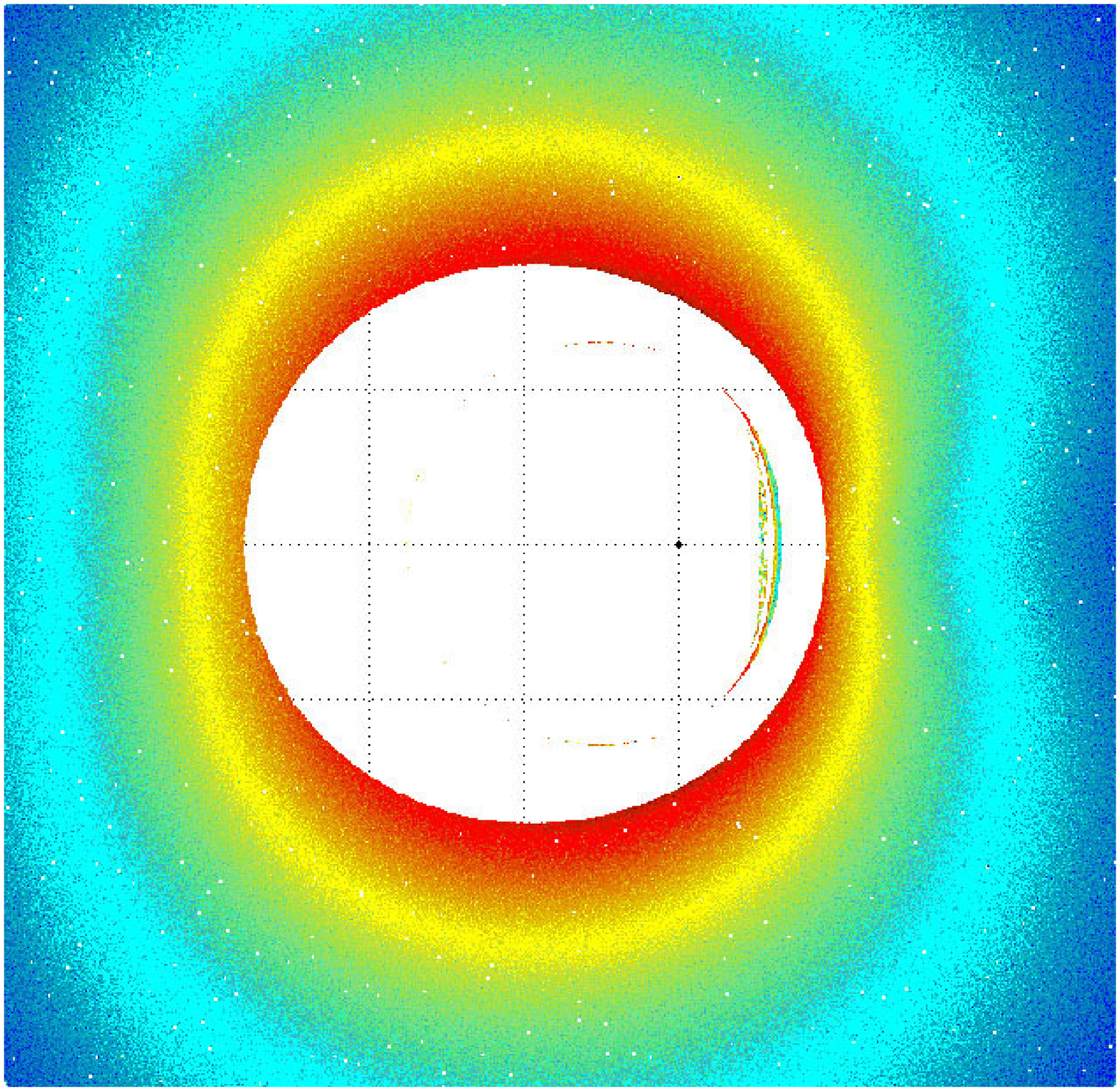}
  \includegraphics[width=4.4cm]{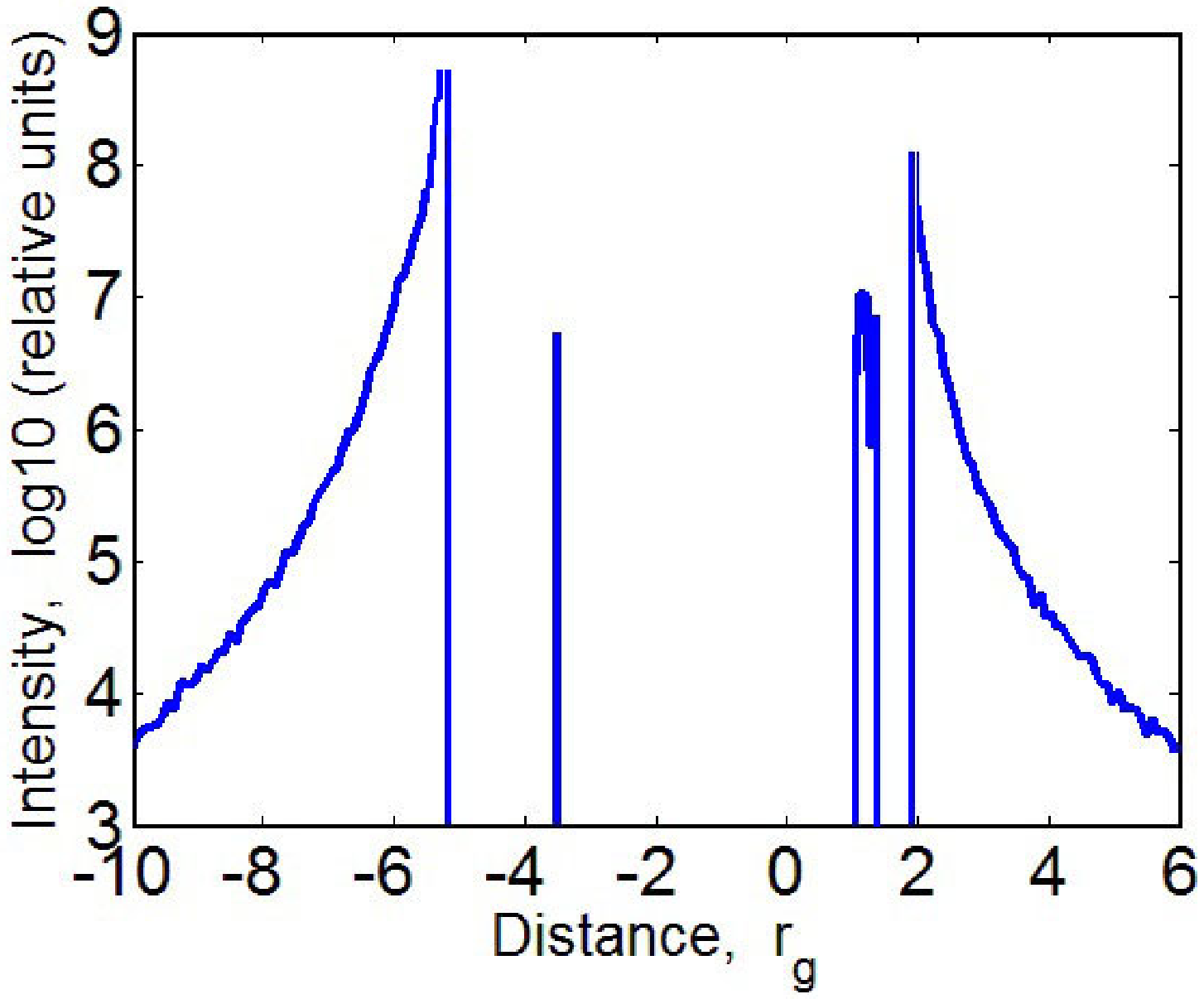}
              }
\centerline{
  \includegraphics[width=4.4cm]{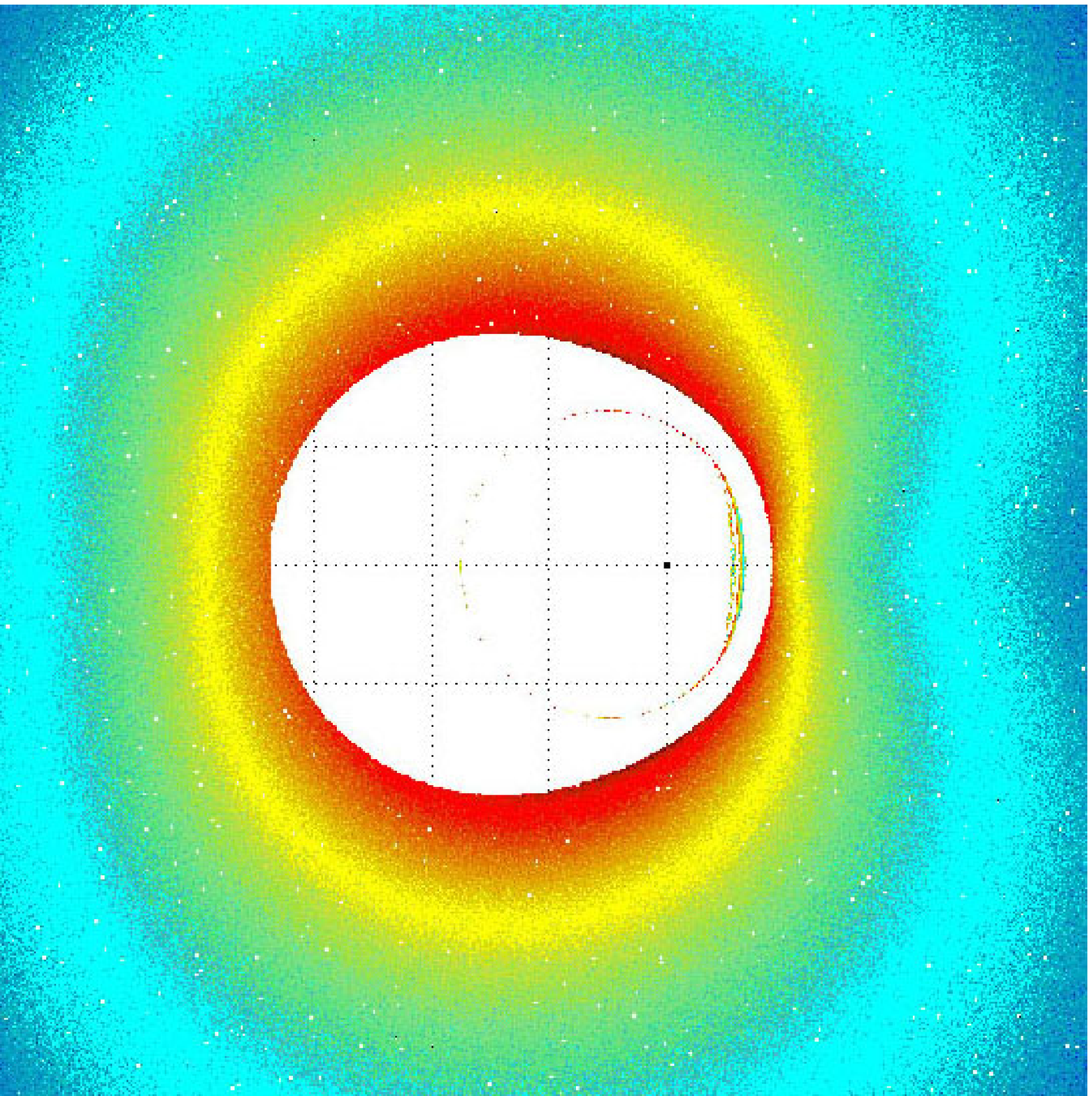}
  \includegraphics[width=4.4cm]{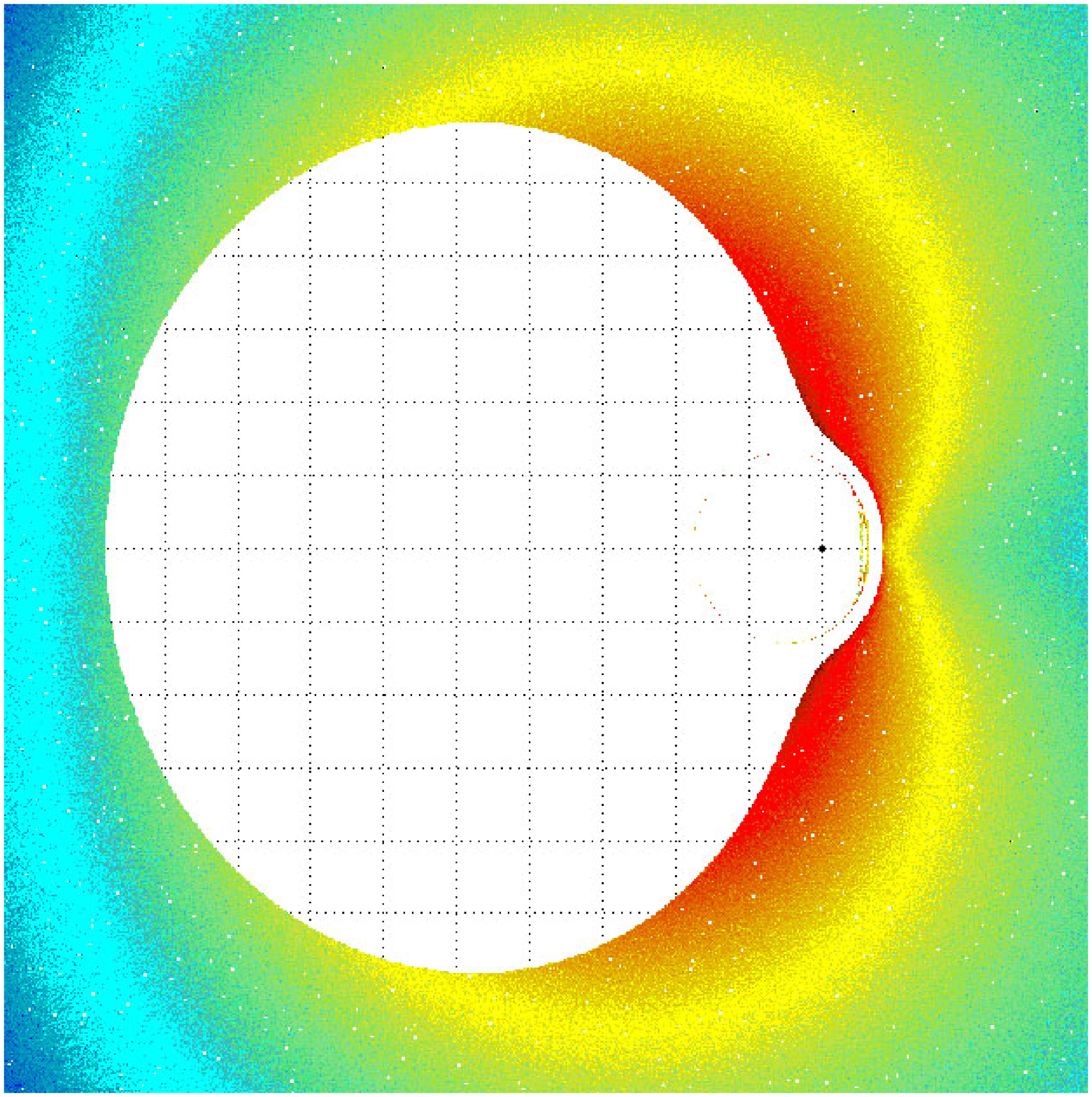}
              }
  \caption{Distribution of the emission intensity when the
           standard screen is on the left side of the observer.
           The intensity is presented in conditional colors;
           the shadow area is white. The grid step is $2r_g$.
           The images correspond to the angles of $45^\circ\!$,
           $60^\circ$ and $80^\circ$ between the view line
           and the standard screen. The top right panel
           demonstrates the distribution of the intensity in
           the equatorial plane for the top left image.}
  \label{Emission_intens2}
\end{figure}

\begin{figure}[tbh]
  \centerline{
  \includegraphics[width=4.4cm]{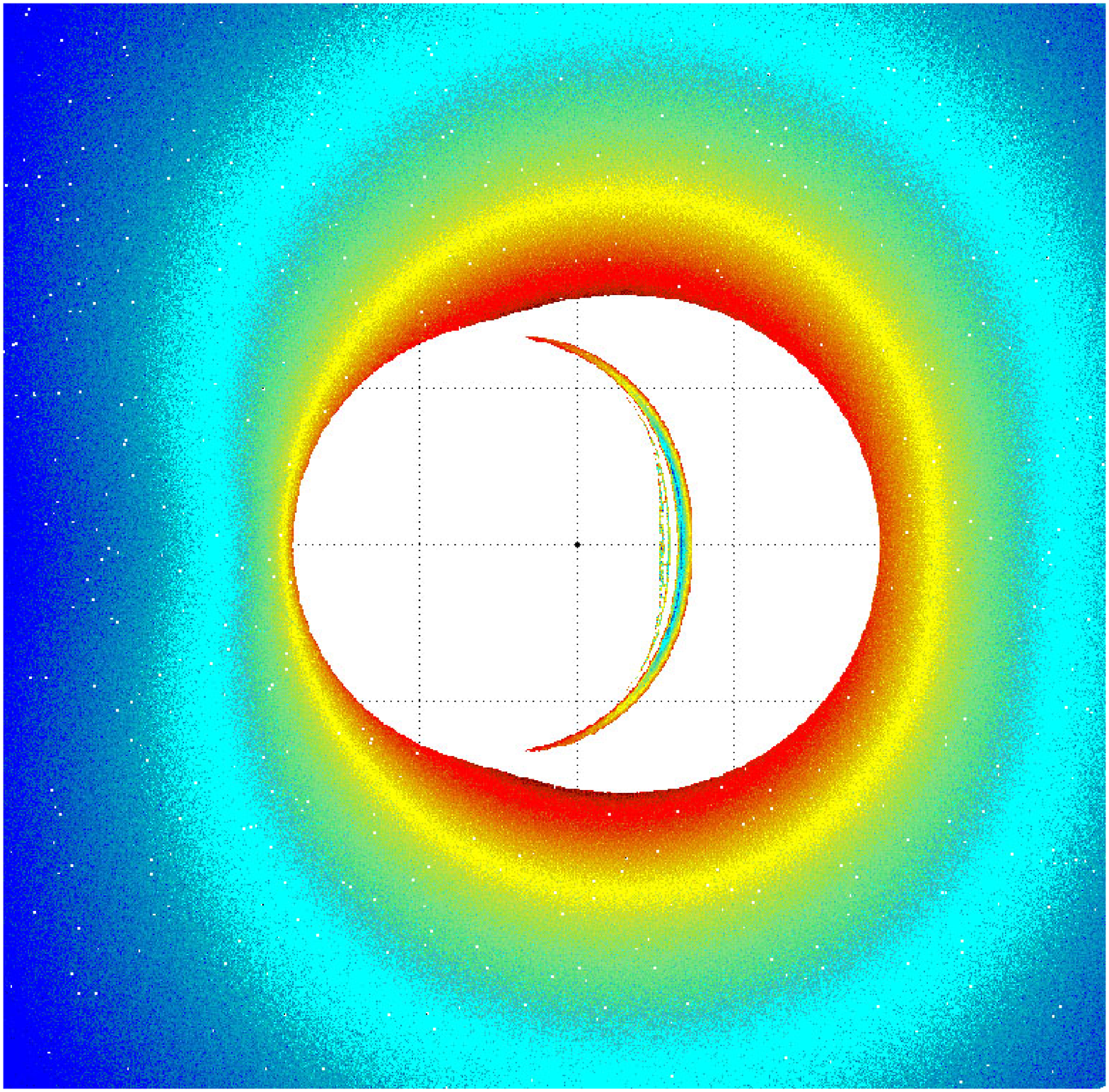}
  \includegraphics[width=4.4cm]{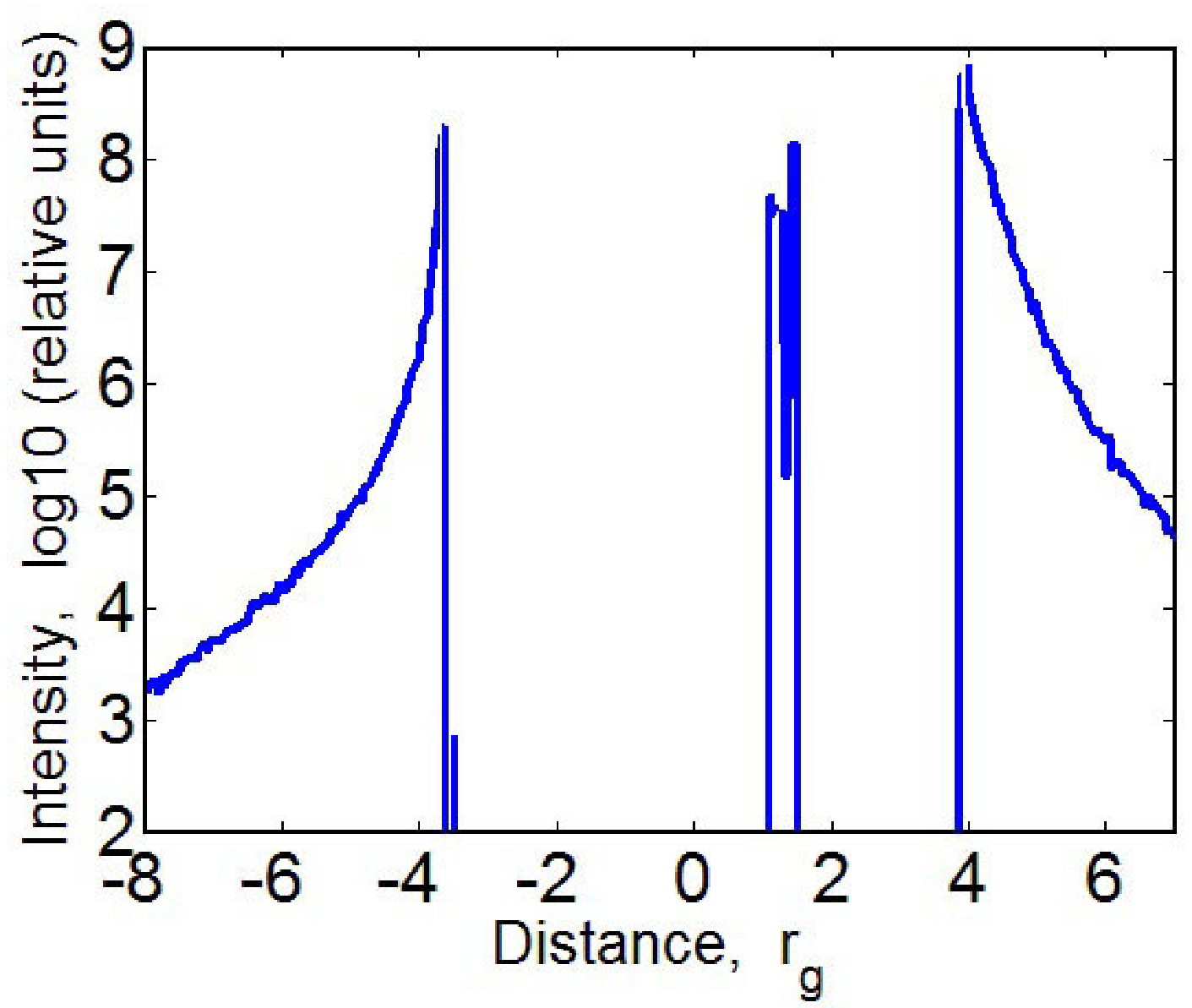}
              }
\centerline{
  \includegraphics[width=4.4cm]{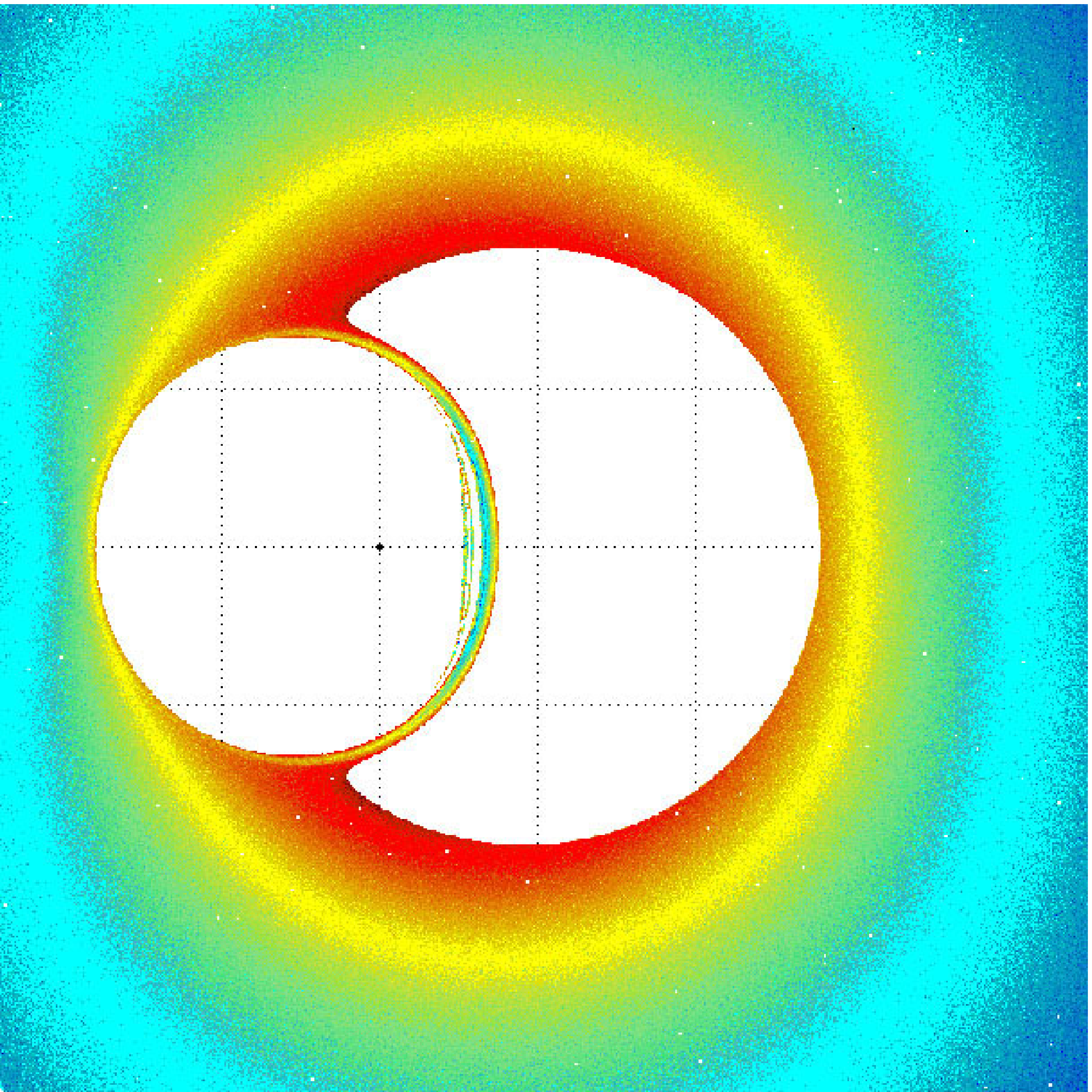}
  \includegraphics[width=4.4cm]{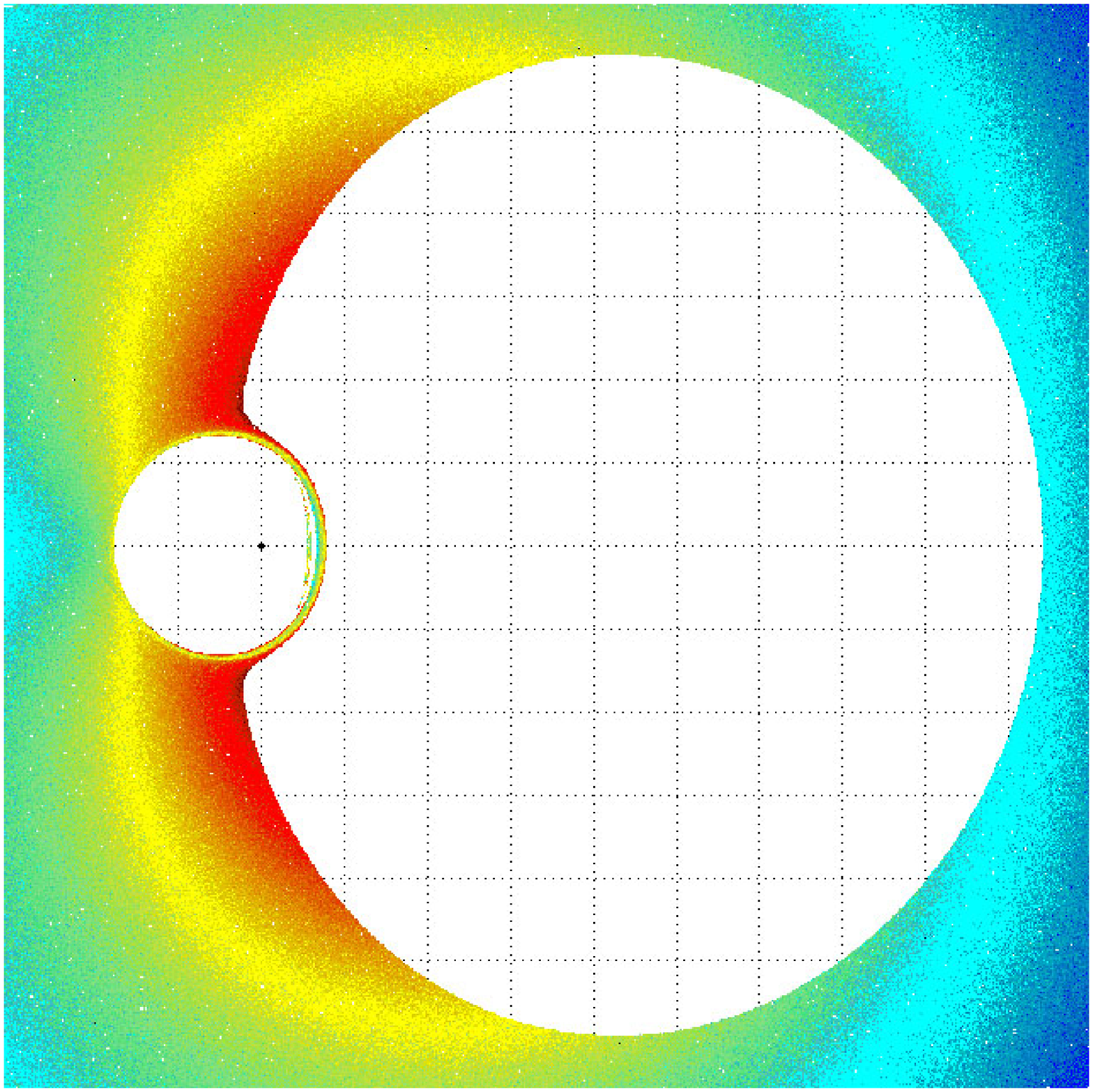}
              }
  \caption{Distribution of the emission intensity when the
           standard screen is on the left side of the observer.
           The intensity is presented in conditional colors;
           the shadow area is white. The grid step is $2r_g$.
           The images correspond to the angles of $-45^\circ\!$,
           $-60^\circ$ and $-80^\circ$ between the view line and
           the standard screen. The top right panel demonstrates
           the distribution of the intensity in the equatorial
           plane for the top left image.}
  \label{Emission_intens3}
\end{figure}

      The image of the shadow when the standard screen is not
perpendicular to the view line is shown in
Figs.~\ref{Emission_intens2}-\ref{Emission_intens3}. The images when
the angle between the view line and the normal to the standard
screen is positive presented in Fig.~\ref{Emission_intens2}.
Conditionally, one can say that the standard screen is located on
the left hand side of the observer. The panels correspond to angles
of $45^\circ$, $60^\circ$ and $80^\circ$. The distribution of the
intensity in the equatorial plane for the angle of $45^\circ$ is
shown on the top right panel. This distribution differs from the one
shown in Fig.~\ref{Emission_intens1}, however the intensity at the
edge of the shadow tends also to infinity.

     The images when the standard screen is to the right side from
the observer, i.e. when the angle between the view line and the
normal to the standard screen is negative is presented in
Fig.~\ref{Emission_intens3}. The panels correspond to the same
angles between the view line and the normal to the standard screen
and the top right panel also shows the intensity in the equatorial
plane for the angle of $45^\circ$. The Figure differs significantly
from its twin. The most interesting point is that the inner annulus
here becomes bold and bright, however its form and size remain
almost unchanged. Moreover, in the annulus we can distinguish even
some very subtle individual rings. At the edges of the shadow the
intens tends to infinity.

\begin{figure}[tbh]
  \centerline{
  \includegraphics[width=4.4cm]{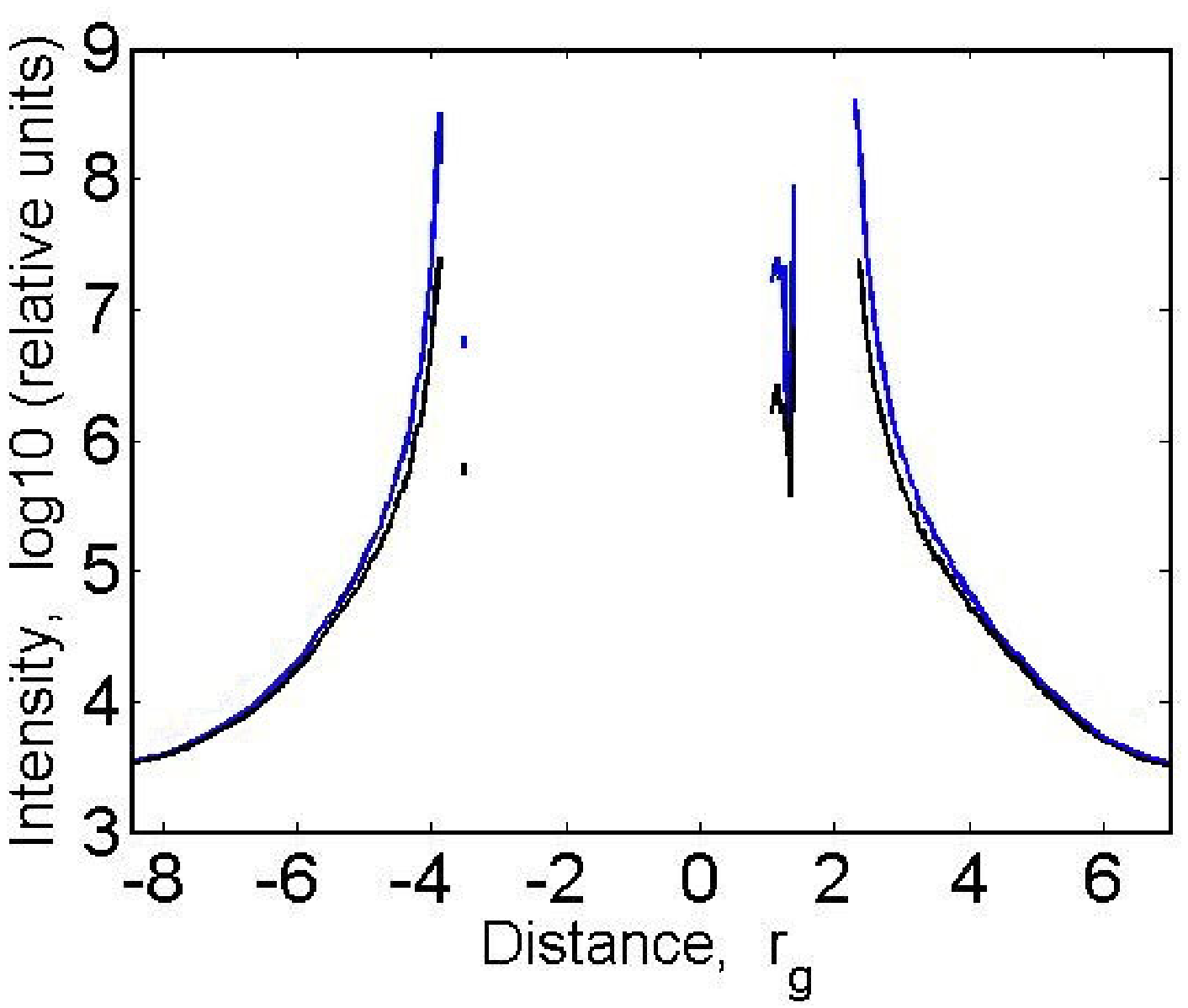}
  \includegraphics[width=4.4cm]{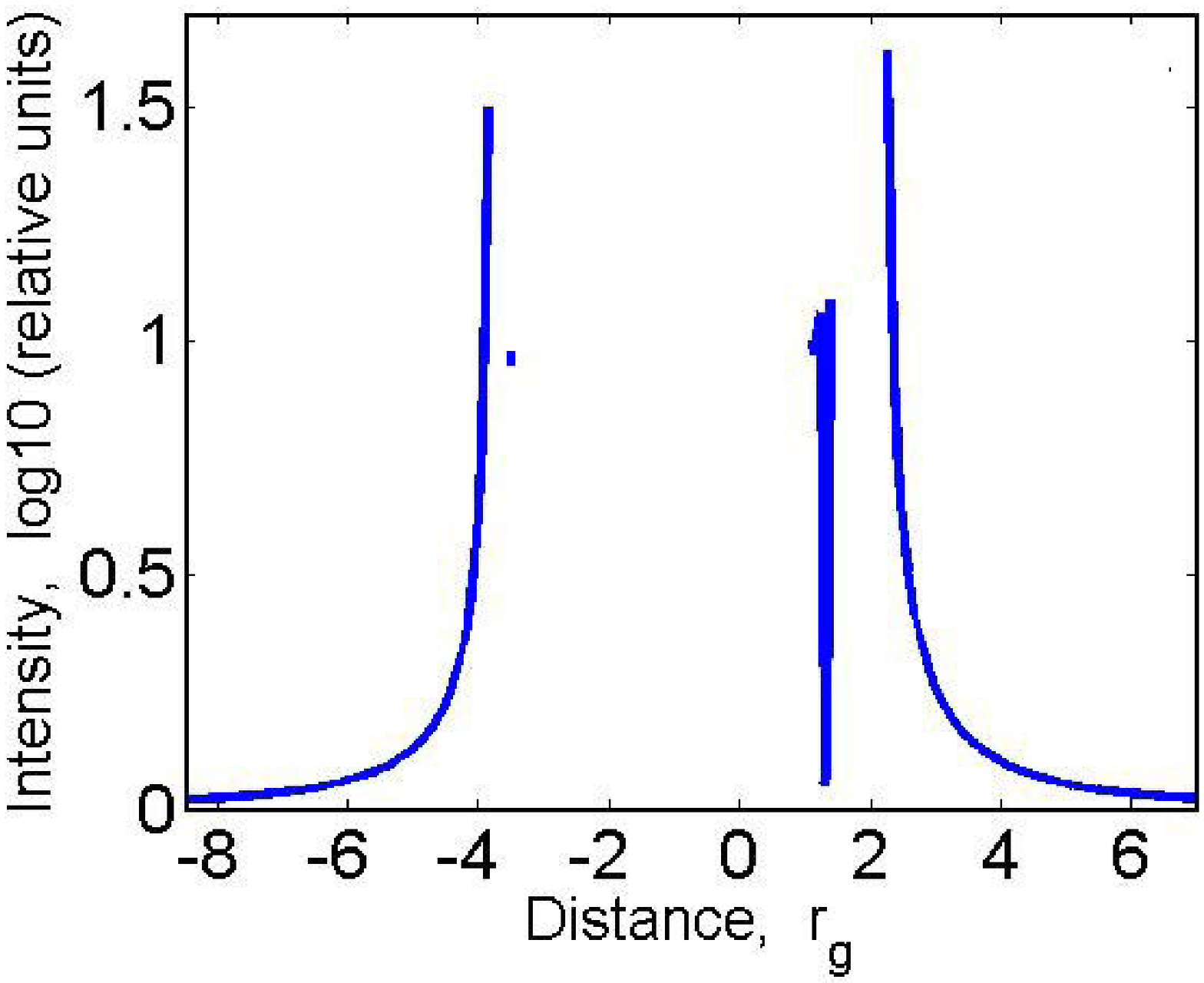}
              }
  \caption{Distribution of the emission intensity for optically
           thin and thick standard screen. One curve repeats the
           plot in Fig.~\ref{Emission_intens1} whereas the other one
           refers to the optically thick screen (the last curve lays
           lower). The right panel demonstrates the difference
           between the two cases in the same scale.}
  \label{Emission_intens4}
\end{figure}

     If the standard screen is optically thick then the distribution
of the emission intensity should differ from the optically thin
case. Both distributions are presented in
Fig.~\ref{Emission_intens4} for the case when the standard screen is
perpendicular to the view line (the distribution for the optically
thin case has already been presented in
Fig.~\ref{Emission_intens1}). As one can see, the difference between
the two curves is so small that it becomes difficult to distinguish
them by the naked eye. The right panel demonstrates this difference
in the same scale. As it follows from the Figures the difference
increases sharply and becomes significant only near the very edge of
the shadow (including the edges of the inner annuli). It means that
the main contribution to the brightness of the ring around the
shadow delivers very far areas of the standard screen.

\section{Discussion}

     The immediate observations of black holes are very
important astrophysical problem, which is not solved yet. One
possible solution is the attempt to observe the black hole shadow,
which can be realized by the interferometer.

     Even in the simplest formulation of the problem, which we
consider, the shadow image appears rather complicated. It depends
not only on the parameters of the black hole, but also on the
characteristics of its surrounding. The real image can be extremely
complicated, so the problem of the extraction of the parameters of
the black hole becomes very difficult, if at all possible, to solve.
Nevertheless, some peculiarities deliver us a hope that in some
cases the problem can be formulated and solved. One of these cases
is the characteristic distribution of the intensity around the
shadow and the appearance of the internal arcs. In the nearest
galaxies, such as M87, Andromeda and our Galaxy, the width of the
arc inside the shadow area may reach 3-4 micro arc seconds. This
detail is bright enough and probably can be registered in the
observations.

     We believe that it is important to separate the peculiarities
of  the shape of the shadow and distribution of the emission
intensity related directly with the black hole from many complicated
factors, such as the presence of the accreting disk with different
distributions of the velocity and the temperature and so on. This
separation is done in this paper.

\section{Acknowledgements}

    One of the authors (S.R.) expresses his gratitude to Dr.
O.Sumenkova, Dr. R.Beresneva and Dr. O.Kosareva for the opportunity
for active and fruitful work on the problem.

    The work has been partly supported by Grant of the President of
the Russian Federation for Support of the Leading Scientific Schools
NSh-6595.2016.2, grant RFBR No. 15-02-00554 and Basic Research
Program P-7 of the Presidium of the Russian Academy of Sciences.

\end{document}